\documentclass{aa}

\usepackage{graphicx}
\usepackage{txfonts}

\DeclareGraphicsExtensions{.eps}

\begin{document}


\title{Exploring the origin of clumpy dust clouds around cool giants}
\subtitle{A global 3D RHD model of a dust-forming M-type AGB star}  

\titlerunning{Clumpy dust clouds around AGB stars} 

\author{
  Susanne H{\"o}fner
    \and
  Bernd Freytag
}

\institute{Theoretical Astrophysics,
           Department of Physics and Astronomy,
           Uppsala University,
           Box 516,
           SE-751 20 Uppsala,
           Sweden \\
           \email{Susanne.Hoefner@physics.uu.se}
}

\date{Received 7 December 2018 / Accepted 31 January 2019}


\abstract
  {Dust grains forming in the extended atmospheres of AGB stars are critical for the heavy mass loss of these cool luminous giants, as they provide radiative acceleration for the stellar winds.
Characteristic mid-IR spectral features indicate that the grains consist mainly of silicates and corundum. The latter species seems to form in a narrow zone within about 2\,stellar radii, preceding the condensation of silicate dust, which triggers the outflow. Recent high-angular-resolution observations show clumpy, variable dust clouds at these distances. 
}
  {We explore possible causes for the formation of inhomogeneous dust layers, using 3D dynamical simulations. 
}
  {We modeled the outer convective envelope and the dust-forming atmosphere of an M-type AGB star with the CO5BOLD radiation-hydrodynamics code. The simulations account for frequency-dependent gas opacities, and include a time-dependent description of grain growth and evaporation for corundum (Al$_2$O$_3$) and olivine-type silicates (Mg$_2$SiO$_4$). 
}
  {In the inner, gravitationally bound, and corundum-dominated layers of the circumstellar envelope,
a patchy distribution of the dust emerges naturally, due to atmospheric shock waves that are generated by large-scale convective flows and pulsations. The formation of silicate dust at somewhat larger distances probably indicates the outer limit of the gravitationally bound layers. The current models do not describe wind acceleration, but the cloud formation mechanism should also work for stars with outflows. Timescales of atmospheric dynamics and grain growth are similar to observed values. In spherical averages of dust densities, more easily comparable to unresolved observations and 1D~models, the variable 3D morphology manifests itself as cycle-to-cycle variations. 
}
 {Grain growth in the wake of large-scale non-spherical shock waves, generated by convection and pulsations, is a likely mechanism for producing the observed clumpy dust clouds, and for explaining their physical and dynamical properties.
}

\keywords{
  convection --
  shock waves --
  stars: AGB and post-AGB --
  stars: atmospheres --
  stars: oscillations (including pulsations) --
  circumstellar matter
}

\maketitle

\section{Introduction}\label{s:intro}

Asymptotic giant branch stars are sources of slow, massive winds, which affect the observable properties and final fate of these cool luminous giants. The outflows are commonly assumed to be driven by radiative pressure on dust grains, that form in the extended stellar atmospheres \citep[for a review on mass loss of AGB stars see, e.g.,][]{Hoefner2018A&ARv..26....1H}. In recent years, the on-going progress in high-angular-resolution techniques has given increasingly detailed insights into the layers of the dynamical atmospheres where the dust grains form and where the outflows have their origin. Spatially resolved observations of nearby AGB stars span wavelengths from the visual and near-IR, which give indications about grain sizes \citep[][]{Norris2012Natur.484..220N, Ohnaka2016A&A...589A..91O, Ohnaka2017A&A...597A..20O}, to the mid-IR region, where lattice modes make it possible to identify the chemical composition of the dust particles \cite[e.g.,][]{Zhao-Geisler2012A&A...545A..56Z, Karovicova2013A&A...560A..75K, Khouri2015A&A...577A.114K}, and the sub-mm regime, where the gas kinematics and temperatures in the circumstellar environment can be traced \citep[e.g.,][]{Ramstedt2014A&A...570L..14R, Vlemmings2017NatAs...1..848V, Khouri2016MNRAS.463L..74K, Khouri2018A&A...620A..75K}. Such observations provide critical constraints on the mass loss mechanism and the unprecedented possibility of testing models of convection and pulsations, atmospheric dynamics, dust formation, and wind acceleration. 

Dust-forming dynamical atmospheres and dust-driven winds of AGB stars have mostly been studied with 1D hydrodynamical models \cite[e.g.,][]{Winters2000A&A...361..641W, Wachter2002A&A...384..452W, Hoefner2003A&A...399..589H, Jeong2003A&A...407..191J, Hoefner2008A&A...491L...1H, Mattsson2010A&A...509A..14M, Bladh2015A&A...575A.105B, Hoefner2016A&A...594A.108H}. Assuming spherically symmetric flows, such simulations describe the varying radial profiles of densities, temperatures, velocities, and dust properties, accounting for shock waves, which are triggered by pulsations and propagate outward through the atmospheres. 

Earlier, these 1D wind models tended to focus on C-type AGB stars (with C/O $>1$ in the atmosphere, due to dredge-up of newly-produced carbon from the interior), where amorphous carbon grains are the main wind-driving dust species. The resulting mass loss rates, wind velocities, spectral energy distributions, and photometric variations are in good agreement with observations \citep[e.g.,][]{Nowotny2011A&A...529A.129N, Nowotny2013A&A...552A..20N, Eriksson2014A&A...566A..95E}. However, due to the strong absorption by carbon dust, the stellar photospheres and inner atmospheres are often obscured, and detailed comparisons with spatially resolved observations can be difficult \cite[e.g.,][]{Paladini2009A&A...501.1073P, Stewart2016MNRAS.455.3102S, Sacuto2011A&A...525A..42S, Wittkowski2017A&A...601A...3W}. 

The dusty envelopes of M-type AGB stars (C/O $<1$) tend to be more transparent in the visual and near-IR regime, giving a better view of the innermost dust-forming atmospheric layers, and allowing for more detailed tests of dynamical atmosphere and wind models \citep[e.g.,][]{Sacuto2013A&A...551A..72S, Aronson2017A&A...603A.116A, Bladh2017A&A...607A..27B}. The latest generation of DARWIN models for M-type AGB stars produces visual and near-IR spectra, light curves, variations of photometric colors with pulsation phase, and wind properties that are in good agreement with observations \citep{Bladh2013A&A...553A..20B, Bladh2015A&A...575A.105B, Hoefner2016A&A...594A.108H}, supporting a scenario where the radiative pressure, which triggers the outflows, is caused by photon scattering on Fe-free silicate grains \citep[][]{Hoefner2008A&A...491L...1H}. Such particles are highly transparent at visual and NIR wavelengths, which results in significantly less radiative heating and smaller condensation distances than for Fe-bearing silicate grains \citep[see, e.g.,][]{Woitke2006A&A...460L...9W, Bladh2012A&A...546A..76B}. In order to provide sufficient radiative pressure by scattering, however, the grains need to be of a size comparable to the wavelengths where the stellar flux peaks, in other words, grain radii should fall in the range of 0.1 -- 1$\,\mu$m. Earlier, it was considered controversial if such large grains could form in the close vicinity of AGB stars. In recent years, however, several observational studies have found dust grains with radii of about 0.1 -- 0.5$\,\mu$m, at distances below 2-3 stellar radii \citep[][]{Norris2012Natur.484..220N, Ohnaka2016A&A...589A..91O, Ohnaka2017A&A...597A..20O}. 

Observations of scattered stellar light in the visual and near-IR regime can give indications of grain sizes, but they provide only indirect constraints on the chemical composition of the dust particles. Mid-IR spectro-interferometric studies of characteristic dust features, on the other hand, suggest that the formation of silicate dust may be preceded by the condensation of corundum, forming a thin gravitationally bound dust layer close to the stellar photosphere \cite[e.g.,][]{Zhao-Geisler2012A&A...545A..56Z, Karovicova2013A&A...560A..75K, Khouri2015A&A...577A.114K}. The radiative pressure on corundum grains is insufficient to cause noticeable dynamical effects, but condensation of silicate mantles on top of corundum cores may speed up grain growth to the critical size regime, and make wind acceleration more efficient \citep[][]{Hoefner2016A&A...594A.108H}. 

Recent high-resolution imaging of nearby AGB stars at visual and infrared wavelengths has revealed complex, non-spherical distributions of gas and dust in the close circumstellar environment \cite[e.g.,][]{Ohnaka2016A&A...589A..91O, Stewart2016MNRAS.457.1410S, Wittkowski2017A&A...601A...3W}. Temporal monitoring shows changes in both atmospheric morphology and grain sizes over the course of weeks or months \cite[e.g.,][]{Khouri2016A&A...591A..70K, Ohnaka2017A&A...597A..20O}. 

Such phenomena cannot be investigated with the 1D atmosphere and wind models mentioned above, which simulate time-dependent radial structures, but assume overall spherical symmetry of the atmosphere and wind. In the 3D ``star-in-a-box'' models by \cite{Freytag2008A&A...483..571F} and \cite{Freytag2017A&A...600A.137F}, on the other hand, an inhomogeneous distribution of atmospheric gas emerges naturally, as a consequence of large-scale convective flows below the photosphere and the resulting network of atmospheric shock waves. The exploratory models of \cite{Freytag2008A&A...483..571F} also indicated that the dynamical patterns in the gas should be imprinted on the dust in the close stellar environment, due to the density- and temperature-sensitivity of the grain growth process. However, these early 3D simulations included amorphous carbon dust, while most of the recent high-angular observations concern M-type AGB stars, which primarily produce silicate and corundum dust with very different micro-physical properties. 

In this paper, we present first global 3D RHD simulations of dust formation in the atmosphere of an M-type AGB star, featuring new physical input and improved numerics compared to our earlier work. We focus on the innermost, gravitationally bound dust layers, mainly consisting of corundum (Al$_2$O$_3$), and use silicate formation to indicate where a radiatively-driven outflow may be triggered, marking the likely outer edge of the gravitationally bound dust shell. In Sect.~\ref{s:setup}, we give a brief overview of the basic physical assumptions and numerical methods. The results are presented in Sect.~\ref{s:results}, and compared to observations in Sect.~\ref{s:discussion}. Finally, a summary and conclusions are given in Sect.~\ref{s:conclusions}.

\section{Setup of global AGB star models}\label{s:setup}

\begin{table*}[htb]
 \begin{center}
  \caption{Basic model parameters and derived quantities 
  \label{t:ModelParam}}
  \begin{tabular}{l|lrrrrrrr|rrrr}
\hline
model & opacities & \!$n_\mathrm{band}$ & $M_\star$ & $M_\mathrm{env}$ & $L_\star$ & \!\!\!$n_x$$\times$$n_y$$\times$$n_z$\! & $x_\mathrm{box}$ & $t_\mathrm{avg}$ & $R_\star$ & $T_\mathrm{eff}$ & $\log g$ \\
 &  &  & $M_\sun$ & $M_\star$ & $L_\sun$ &  & $R_\sun$ & yr & $R_\sun$ & K & (cgs) \\ \hline
 st28gm06n25 &                    {\footnotesize\verb|phoenix_opal_greynodust01|}\!\! &  1 & 1.0 & 0.182 &  6890 & 401$^3$ & 1970 & 23.77 &  372 & 2727 & -0.71 \\
st28gm06n038 &                     {\footnotesize\verb|t2800gm050mm00_coma_opal|}\!\! &  3 & 1.0 & 0.182 &  7049 & 401$^3$ & 1970 & 26.94 &  478 & 2417 & -0.92 \\
st28gm06n039 &                     {\footnotesize\verb|t2800gm050mm00_coma_opal|}\!\! &  1 & 1.0 & 0.182 &  7027 & 401$^3$ & 1970 & 26.98 &  386 & 2690 & -0.74 \\
\hline
  \end{tabular}
 \end{center}
{The table shows the model name;
the name of the opacity file;  
the number $n_\mathrm{band}$ of frequency bands or bins, used in the radiation transport (1\,=\,gray);
the mass $M_\star$, used for the external potential;
the envelope mass $M_\mathrm{env}$, derived from integrating the mass density of all grid cells within the computational box; 
the average emitted luminosity $L_\star$; 
the model grid dimensions $n_x$$\times$$n_y$$\times$$n_z$;
the edge length of the cubical computational box $x_\mathrm{box}$;
the time $t_\mathrm{avg}$, used for averaging the remaining quantities in this table; 
the average approximate stellar radius $R_\star$;
the average approximate effective temperature $T_\mathrm{eff}$;
the logarithm of the average approximate surface gravity $\log g$.
The pulsation period of model st28gm06n25 is 1.388 years (507 days), the other models have similar periods. 
}
\end{table*}

Below, we give a short summary of the physical and numerical properties of the CO5BOLD code, relevant for the new simulations presented in this paper. More details can be found in our earlier papers on global AGB star models \citep{Freytag2008A&A...483..571F, Freytag2017A&A...600A.137F}. The newly-implemented routines describing the condensation of corundum and silicate grains are based on the dust model discussed in \cite{Hoefner2016A&A...594A.108H}.

\subsection{Properties of the CO5BOLD code}

The CO5BOLD code \citep{Freytag2012JCP...231..919F} numerically integrates the coupled non-linear equations of compressible hydrodynamics and non-local radiative energy transfer. The hydrodynamics scheme is based on an approximate Riemann solver of Roe-type \citep[see also][]{Freytag2013MSAIS..24...26F}, modified to account for the effects of ionization and gravity. The non-local radiative energy transfer in the global models is solved with a short-characteristics scheme.
The numerical grid is Cartesian. In all models presented here, the computational domain and all individual grid cells are cubical. All outer boundaries are open for the flow of matter and for radiation
\citep[see][for some details about boundary conditions in CO5BOLD]{Freytag2017MmSAI..88...12F}.

Gravitation is included as an external potential, with a general $1/r$ profile, that is smoothed in the central region of the star. In this volume, heat is added as a constant source term to produce the desired stellar luminosity, since the tiny central region where nuclear reactions take place cannot possibly be resolved with grid cells of constant size. A drag force is active in this core region only,
to prevent dipolar flows traversing the entire star. The tabulated equation of state (assuming solar abundances) takes the ionization of hydrogen and helium and the formation of H$_2$ molecules into account.

Our earlier global 3D RHD simulations of AGB stars used tabulated gray opacities merged from Phoenix \citep{Hauschildt1997ApJ...483..390H} and OPAL \citep{Iglesias1992ApJ...397..717I} data
(see Table\,\ref{t:ModelParam}). This is sufficient for studies of interior properties like convection and pulsations and to give a qualitative picture of the shock-dominated atmospheric dynamics \citep[see][]{Freytag2017A&A...600A.137F}. The new dust-forming simulations discussed in this paper, on the other hand, require a refined modeling of the atmospheric temperature structure
using a frequency-dependent table. \citet[][]{Chiavassa2011A&A...535A..22C} presented such non-gray models for the case of red supergiants. The frequency dependence of the opacities is treated in an approximate way by an opacity-binning technique based on a method presented by \citep{Nordlund1982A&A...107....1N} and refined later on \citep[see][and references therein]{Freytag2012JCP...231..919F}. The binning scheme for the new tables does not use a prescribed set of sorting rules \citep[as, e.g., in][]{Nordlund1982A&A...107....1N}. Instead, it employs an iterative procedure to minimize the error in the overall heating and cooling rate for a representative one-dimensional temperature-pressure stratification. The error is computed from the differences between solutions of the radiative transfer equation based on the binned and on the full opacities. The new tables for atmospheric gas opacities are based on COMA data \citep[see][]{Aringer2000DissAri, Aringer2016MNRAS.457.3611A} extended with OPAL data at temperatures above approximately 12\,000\,K. Scattering is treated as true absorption. Dust opacities and radiation pressure are not taken into account at present (see below).

\subsection{Dust species: corundum and silicates}\label{s:dust_spec}

To account for the effects of dynamical processes in the stellar atmosphere on dust formation, 
we use a time-dependent kinetic treatment of grain growth, as described in detail by \cite{Hoefner2016A&A...594A.108H}. The grains grow by addition of abundant atoms and molecules from the gas phase, and they may shrink due to thermal evaporation from the grain surface. Corundum (Al$_2$O$_3$) is assumed to form according to the net reaction
\begin{equation}
  {\rm 2 \, Al + 3 \, H_2 O }  \,\,  \longrightarrow  \,\,  {\rm Al_2 O_3 + 3 \, H_2 } \, .
\end {equation}
Since O is much more abundant than Al in a solar mixture, the grain growth rate and the maximum amount of corundum that can form will be limited by the abundance of Al. The condensation of Fe-free olivine-type silicates (Mg$_2$SiO$_4$) is assumed to proceed according to the net reaction
\begin{equation}\label{e_path_ol}
  {\rm 2 \, Mg + SiO + 3 \, H_2 O }  \,\,  \longrightarrow  \,\,  {\rm Mg_2 SiO_4 + 3 \, H_2 } \, .  \end{equation}
In a solar element mixture, the abundance of Si and Mg are comparable, and the abundance of SiO will be determined by the abundance of Si in the gas phase. Since two Mg atoms are required for each SiO molecule added to the dust particles, Mg will be the limiting factor for grain growth under these circumstances. Due to less efficient radiative heating (lower NIR absorption) Fe-free silicates can exist closer to the star than their Fe-bearing counterparts, thereby marking the inner edge of the silicate formation zone.

The kinetic treatment of grain growth summarized above does not describe nucleation, that is, the formation of the very first solid condensation nuclei (often referred to as seed particles) out of the gas phase. Since the chemical formation pathways and nucleation rates for seed particles in M-type AGB stars are still a matter of debate \citep[e.g.,][and references therein]{Gail2016A&A...591A..17G, Gobrecht2016A&A...585A...6G, Kaminski2018arXiv180910583K}, the abundance of seed particles relative to hydrogen is treated as an input parameter, and set to a value of $3 \cdot 10^{-15}$ \citep[see the discussion in][]{Hoefner2016A&A...594A.108H}. It is assumed that these seed particles are readily available whenever conditions permit the condensation of corundum or silicate dust according to the kinetic grain growth scheme outlined above. With a radius of about $2 \cdot 10^{-7}$\,cm (corresponding to 1000 monomers) the seed particles are tiny compared to the resulting dust grains, and they have no effect other than providing an initial condensation surface for grain growth in the context of the models presented here. The growth of grains is triggered by temperature falling below a critical value \citep[see Fig.\,1 in][]{Hoefner2016A&A...594A.108H}, which causes the gas to be supersaturated. Reversely, when the temperature rises above this value, the grains start to shrink due to evaporation from the surface. The critical temperature for a given dust species depends on the prevailing densities, as does the grain growth rate. At the relatively low densities in the stellar atmosphere, grain growth typically takes weeks to months. This is comparable to the timescales of gas dynamics and radiative flux variations, and grain growth may therefore proceed far from equilibrium. 

At present, both corundum and silicate dust are considered as ``passive'' components in the 3D models. In other words, they react to the conditions in the gas, but their opacities are not fed back into the radiative transfer, and their temperatures are set by the gas temperature. Future studies of dust-driven outflows based on global 3D simulations with CO5BOLD will require considerable work on the code. Dust opacities have to be accounted for and a treatment of radiative pressure has to be implemented. In addition, a significantly larger computational domain will be required to include
at least the critical inner part of the wind-driving zone. The development is in progress but not finished, yet. Consequently, radiative pressure is not taken into account in the simulations presented here, and they focus on the innermost, gravitationally bound part of the circumstellar envelope where the dust clouds emerge. This region is dominated by corundum dust, which should not directly affect the local dynamics due to its low flux-mean opacity and negligible radiative pressure \citep[see discussion in][]{Hoefner2016A&A...594A.108H}.

Since corundum can condense at higher temperatures than Mg-Fe silicates, it has been discussed in the literature if these species will occur in composite grains, with a silicate mantle condensing on top of a corundum core formed closer to the star \citep[e.g.,][]{Kozasa1997Ap&SS.255..437K, Kozasa1997Ap&SS.251..165K}. This scenario is consistent with recent findings in meteoritic grains \citep[e.g.,][]{Leitner2018GeCoA.221..255L}. In the models presented here, however, the two types of dust are treated as forming separate grains, each growing on tiny pre-existing seed particles (see above). This allows us to study the complex effects of 3D atmospheric dynamics and a variable radiative flux on dust condensation and evaporation, without the added  complication of the two dust species being interdependent. \citet [][]{Hoefner2016A&A...594A.108H} demonstrated that condensation of silicate mantles on top of corundum cores may speed up grain growth to the critical size regime for effective photon scattering, and make wind acceleration more efficient. In future 3D models which take radiation pressure on dust into account, the possible formation of core-mantle grains should therefore be considered. In the present study, however, a separate treatment of the two dust species will neither affect dynamics, nor the basic cloud formation mechanism, which operates near the inner edge of the dusty envelope, preceding silicate formation.

\begin{figure}[hbtp]
\includegraphics[width=8.8cm]{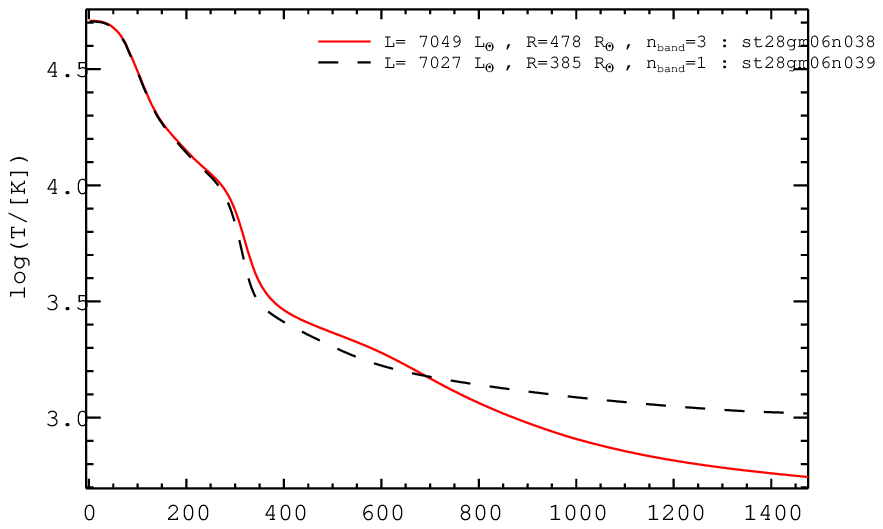}\vspace{-2mm}
\includegraphics[width=8.8cm]{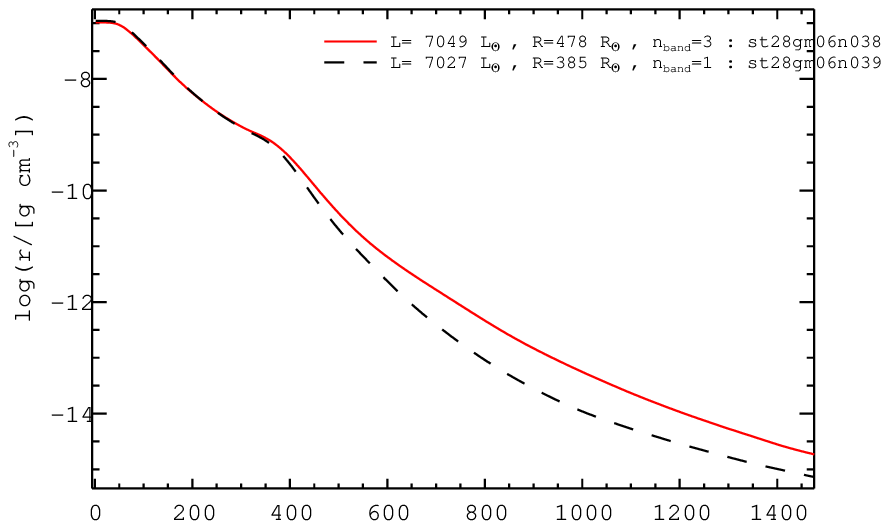}\vspace{-2mm}
\includegraphics[width=8.8cm]{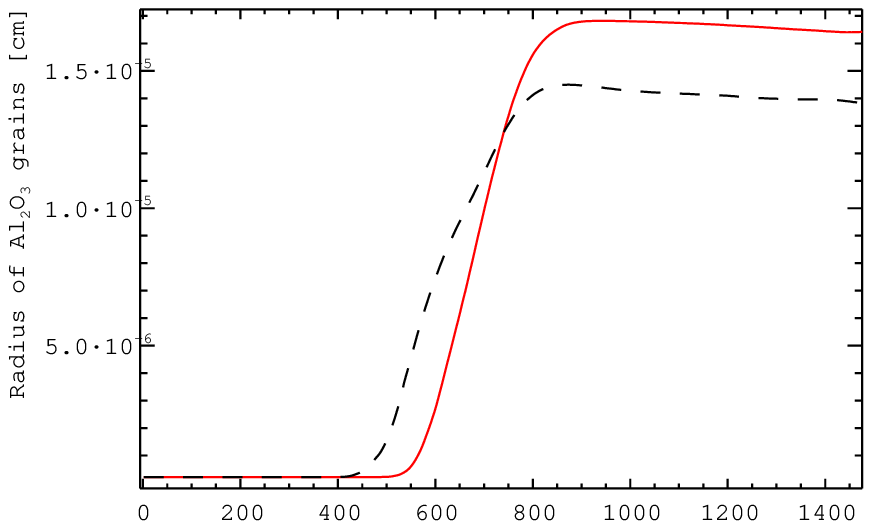}\vspace{-2mm}
\includegraphics[width=8.8cm]{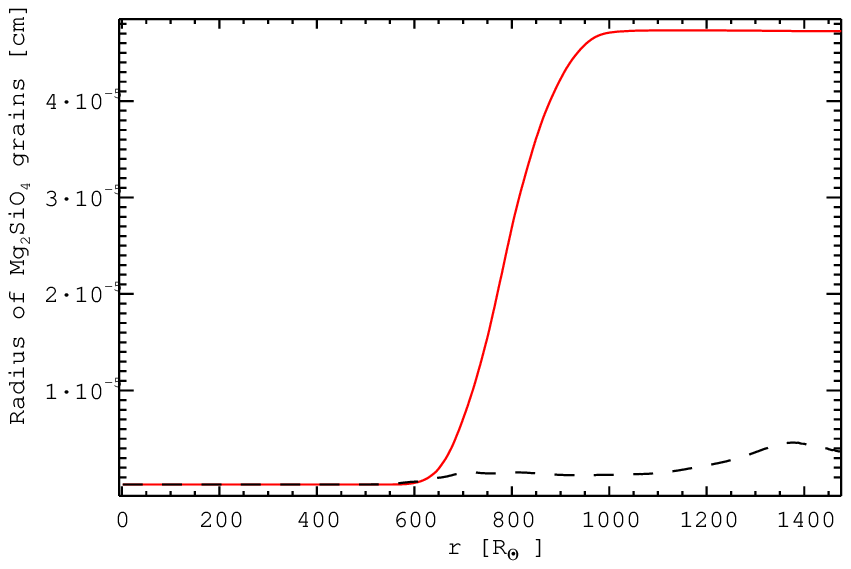}
\caption{Mean structures of the non-gray model st28gm06n038 (continuous red curves)
and the gray model st28gm06n039 (dashed black curves). Shown are 
temperature,
gas density,
corundum grain radius, and
silicate grain radius,
averaged over spherical shells and time, and plotted against the distance from the stellar center. 
\label{f:st28gm06n038_TimeAvgx}}
\end{figure}

\subsection{3D models: Input parameters and resulting quantities}

The new dust-forming 3D models discussed in this paper use a snapshot of the dust-less model st28gm06n25 of \cite{Freytag2017A&A...600A.137F} as a starting point, but they are based on different tabulated gas opacities, as discussed above, in order to take effects of non-gray radiative transfer into account. The properties of the models are summarized in Table\,\ref{t:ModelParam}. While the mass $M_\star$ (controlling the gravitational potential), as well as the resolution and the extent of the numerical grid, are pre-chosen fixed parameters, other model properties are determined after a simulation is finished. The stated stellar luminosity is a time average of the total luminosity emitted at the surface (very close but not identical to the inserted luminosity of 7000\,$L_\sun$ in the core). The envelope mass $M_\mathrm{env}$ is calculated from the integrated density of all grid cells, averaged over time. The radius is more difficult to determine and less well defined due to the complex morphology of the extended atmosphere. It is chosen as the point $R_\star$ where the spherically and temporally averaged temperature and luminosity fulfill $\langle L \rangle_{\Omega,t}$\,=$\,4\pi\sigma R_\star^2 \langle T \rangle_{\Omega,t}^4$.

The parameter $n_\mathrm{band}$ listed in Table\,\ref{t:ModelParam} specifies the number of frequency bins used in the computation of the radiative energy transfer (1\,=\,gray). The new non-gray model st28gm06n038 is characterized by a larger stellar radius and lower effective temperature than its gray parent model st28gm06n25. To check that this is an effect of non-gray radiative transfer, and not caused by the differences in the underlying gas opacity data, we also computed a gray version of the new model, st28gm06n039, based on the same new opacities. As can be seen, the resulting stellar radius and effective temperature are close to the values of the original gray model st28gm06n25, and differ considerably from the corresponding non-gray model st28gm06n038.   

In Fig.\,\ref{f:st28gm06n038_TimeAvgx}, we compare the mean radial structures of the new non-gray and gray models, showing temperatures, densities, and dust grain radii, averaged over spherical shells and time. While the models are almost identical in the convective, optically thick stellar interior
(above $T$\,$\gtrsim$\,8000\,K), they differ noticeably in the atmosphere. The inner atmosphere of the non-gray model st28gm06n038 is slightly hotter than that of its gray counterpart st28gm06n039. However, the models differ strongly in the outer atmospheric layers, where molecular opacities affect the radiative flux. Here, the non-gray model is considerably cooler than the gray model, leading to a much more efficient growth of silicate grains. Corundum, forming closer to the star where the differences are smaller, is less affected, but also shows a higher maximum degree of condensation in the non-gray case. Tests varying the number of frequency bins indicate that three bins are already sufficient to include the essential features of a non-gray stratification, for example, low temperatures in the outer atmosphere and short radiative time scales. A larger number of bins has only a minor impact on the temperature and density stratification, while further increasing the already considerable CPU time.

\begin{figure*}[hbtp]
\begin{center}
\hspace*{0.9cm}\includegraphics[width=15.3cm]{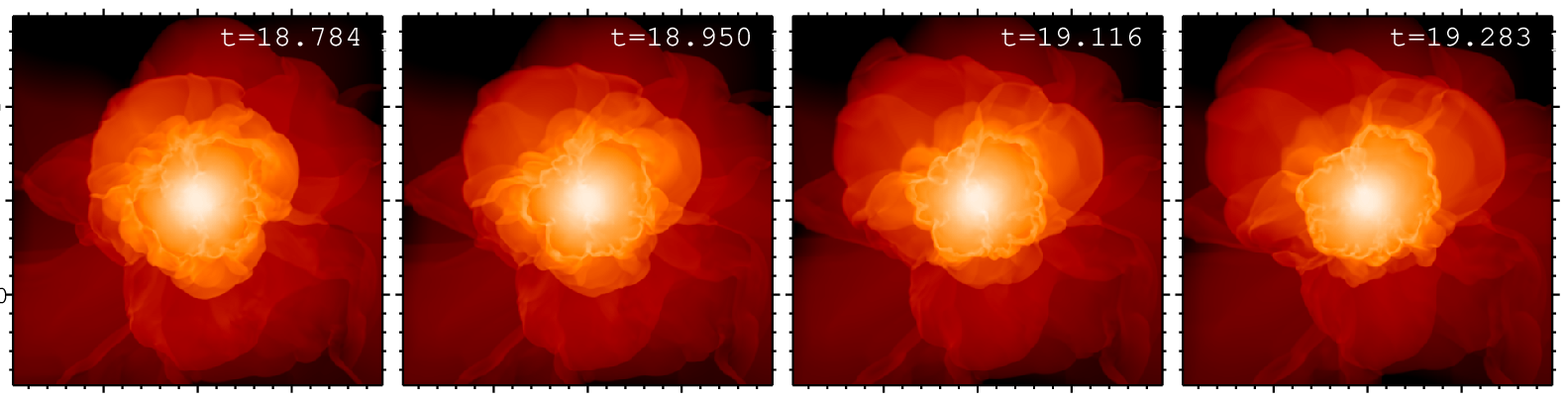}\includegraphics[width=1.9125cm]{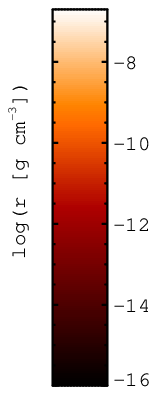}

\hspace*{0.9cm}\includegraphics[width=15.3cm]{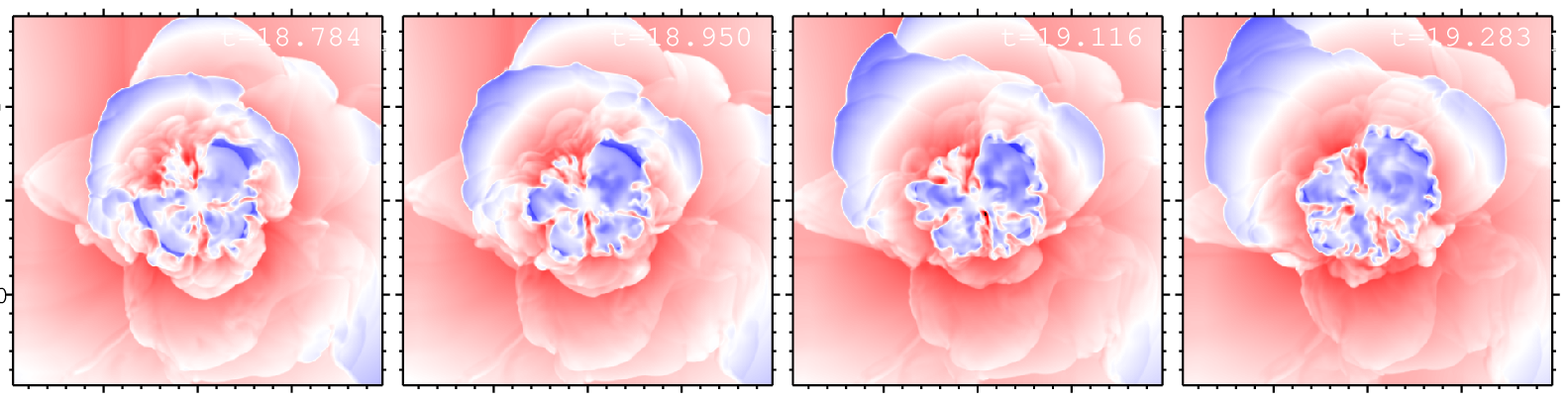}\includegraphics[width=1.9125cm]{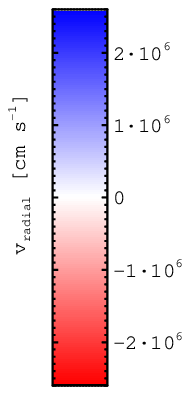}

\hspace*{0.9cm}\includegraphics[width=15.3cm]{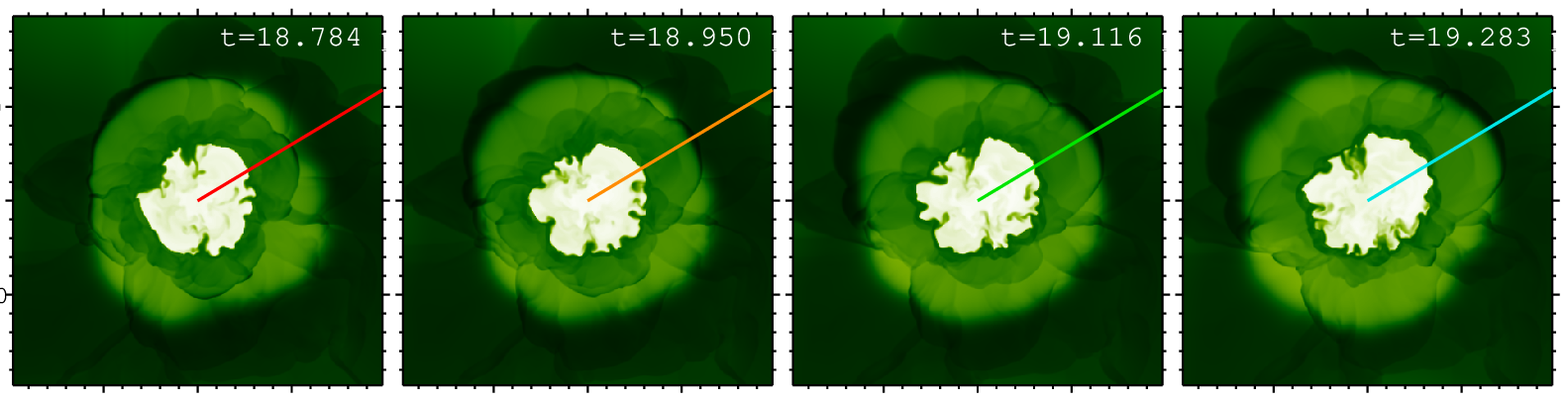}\includegraphics[width=1.9125cm]{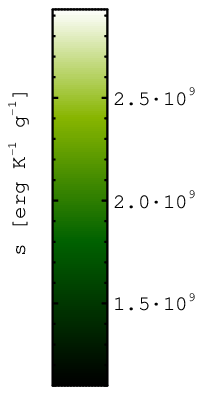}

\hspace*{0.9cm}\includegraphics[width=15.3cm]{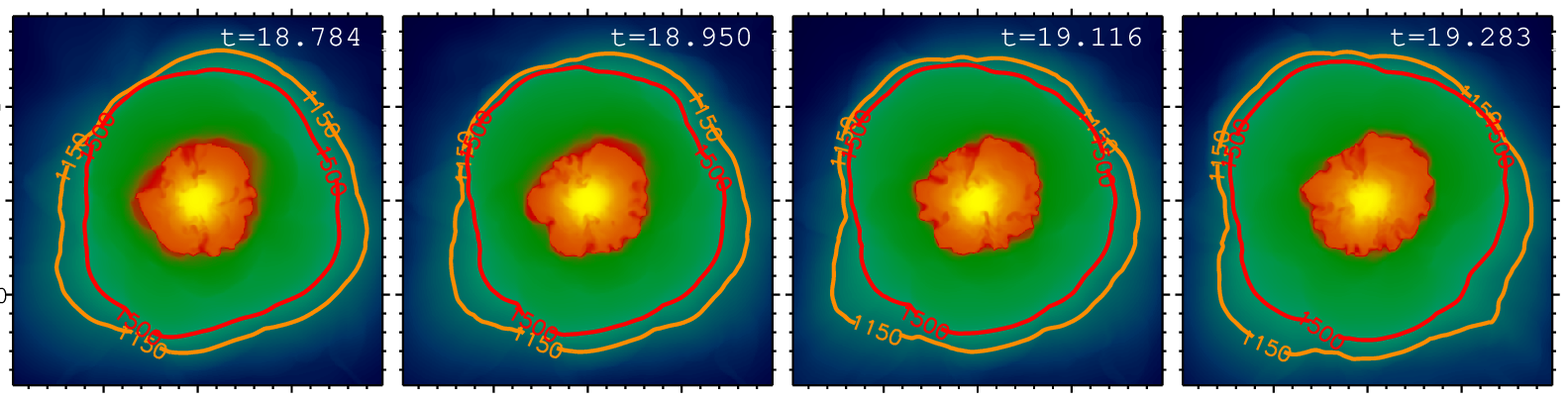}\includegraphics[width=1.9125cm]{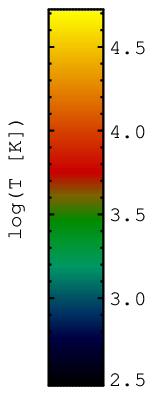}

\hspace*{0.9cm}\includegraphics[width=15.3cm]{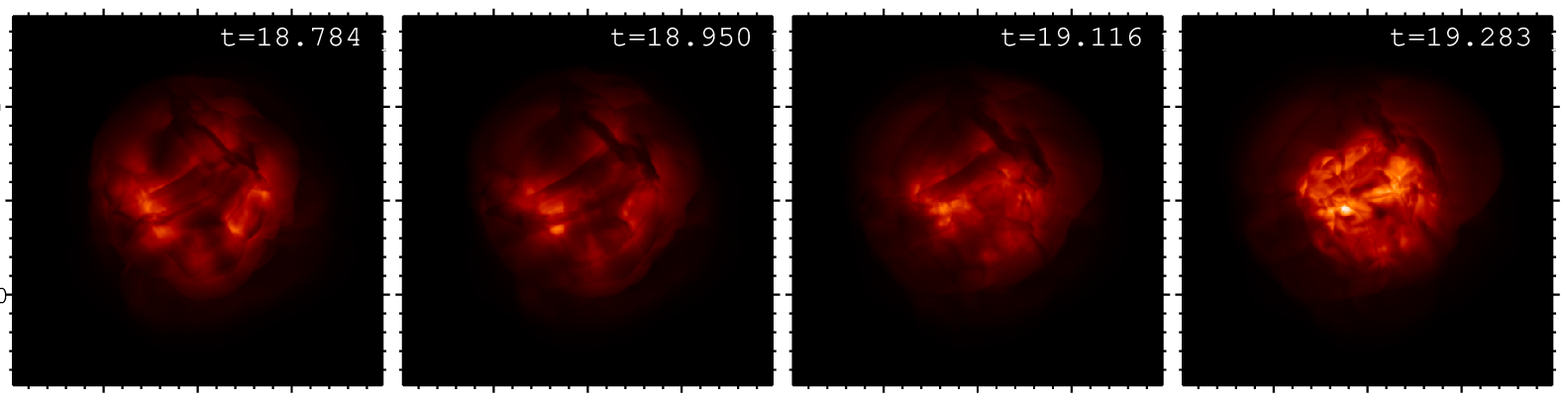}\includegraphics[width=1.9125cm]{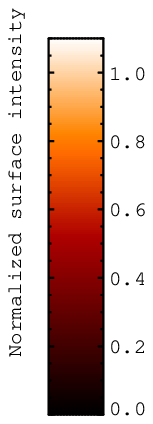}\vspace{0.3cm}
\end{center}
\caption{Time sequences of density, radial velocity, entropy, and temperature for slices through the center of model st28gm06n038 (rows 1--4), and the variation of relative surface intensity (bottom row).
The snapshots are about 2 months apart (see the counter in the top of the panels).
Colored lines in the middle row indicate the radial ray used in Fig.\,\ref{f:st28gm06n038_1DRay}, 
following the emergence of a new dust cloud. The contour lines in the temperature panels at 1500\,K and 1150\,K roughly correspond to the inner edges of the corundum and silicate dust layers, respectively (see Fig.\,\ref{f:st28gm06n038_RdustTandrhoContour}). 
\label{f:st28gm06n038_QuSeq1}}
\end{figure*}

\begin{figure*}[hbtp]
\begin{center}
\hspace*{0.9cm}\includegraphics[width=15.3cm]{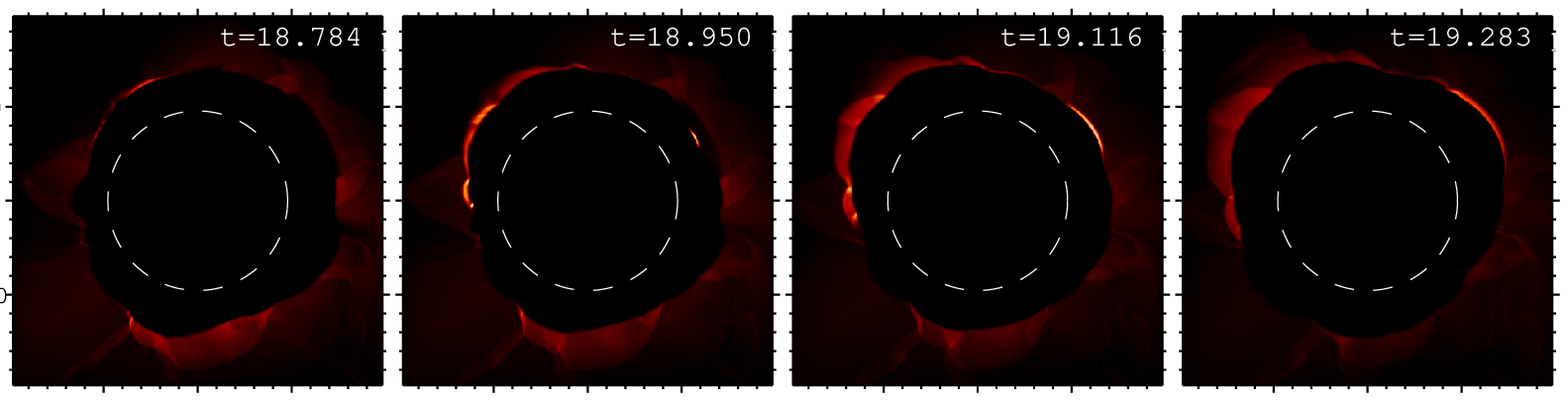}\includegraphics[width=1.9125cm]{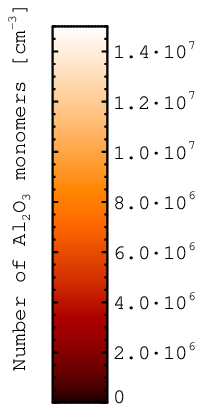}

\hspace*{0.9cm}\includegraphics[width=15.3cm]{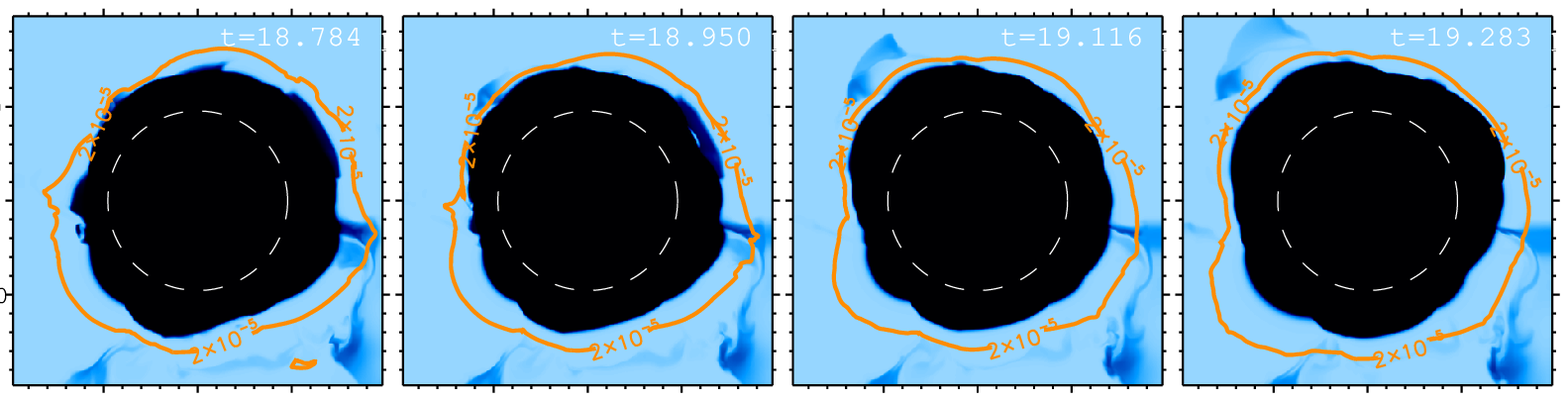}\includegraphics[width=1.9125cm]{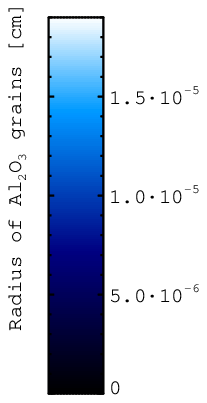}

\hspace*{0.9cm}\includegraphics[width=15.3cm]{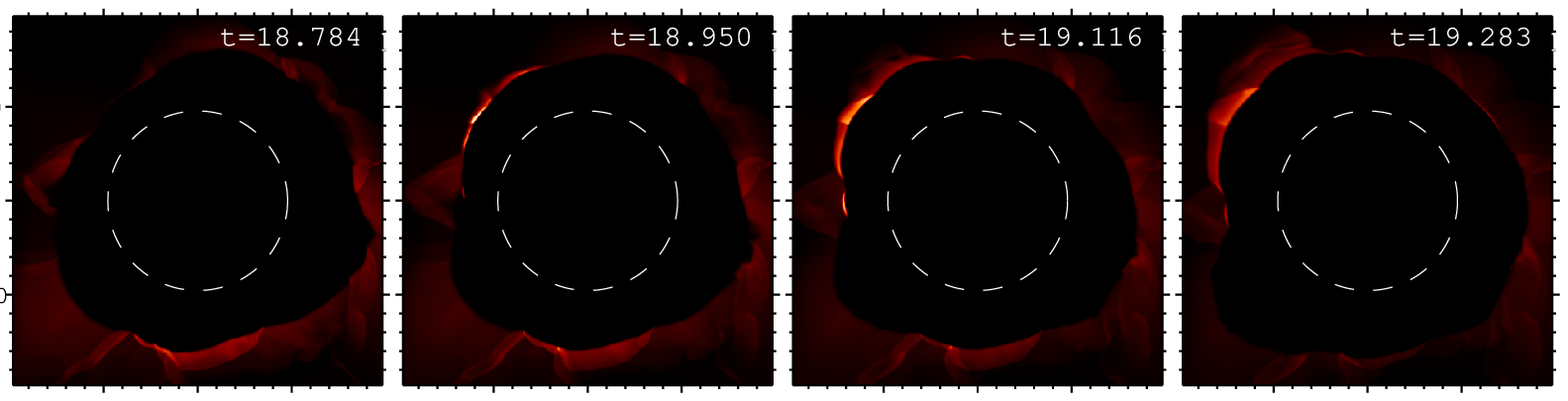}\includegraphics[width=1.9125cm]{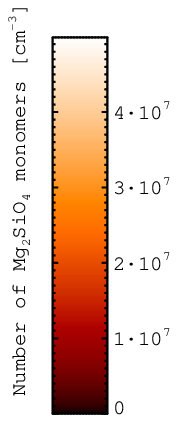}

\hspace*{0.9cm}\includegraphics[width=15.3cm]{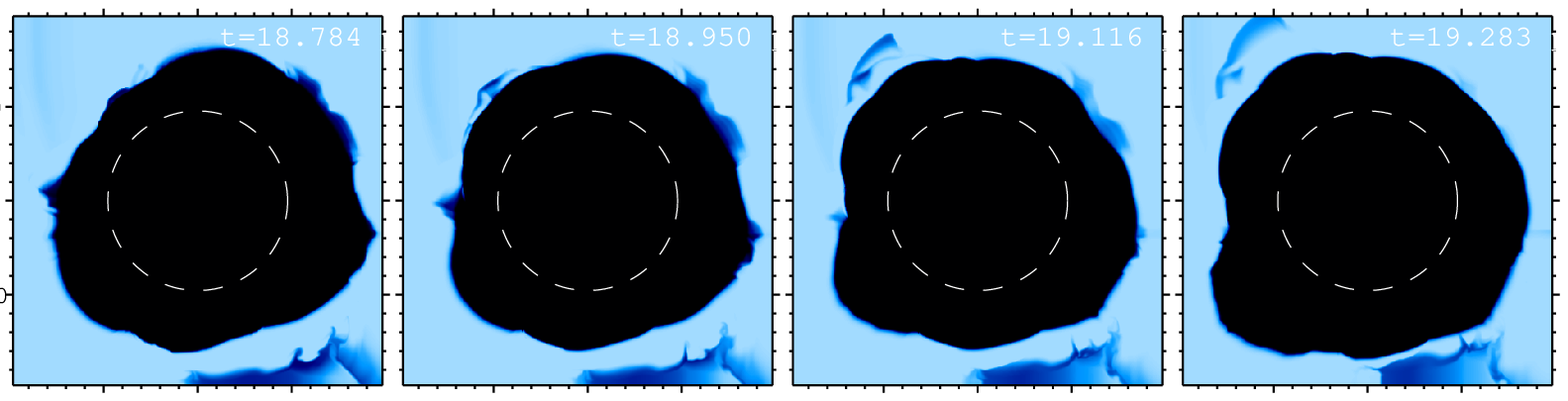}\includegraphics[width=1.9125cm]{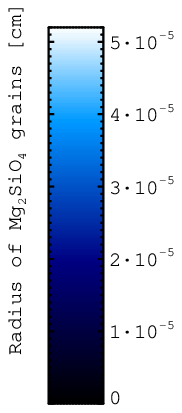}

\hspace*{0.9cm}\includegraphics[width=15.3cm]{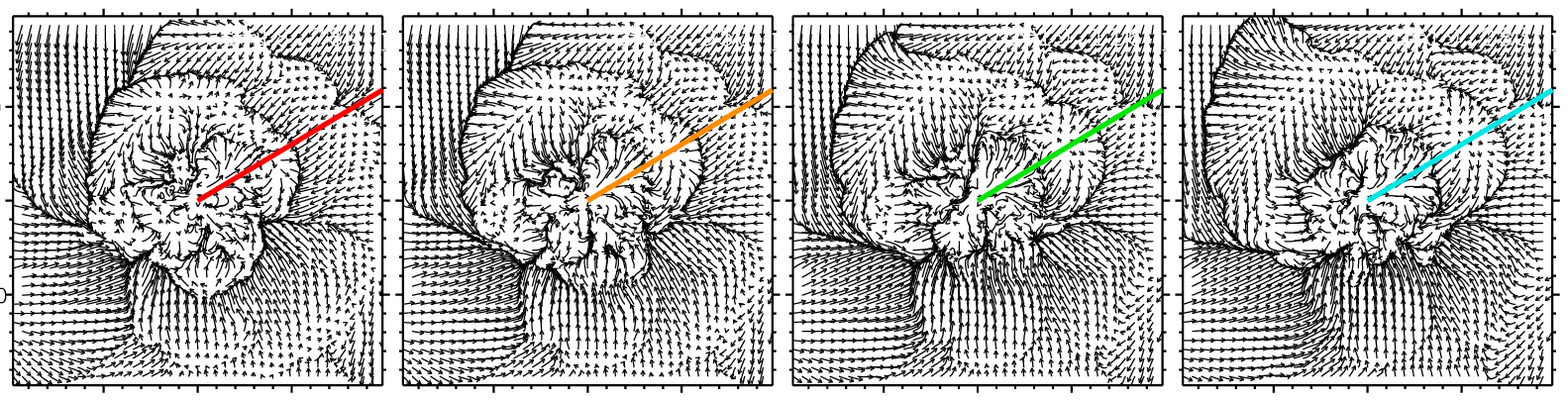}\hspace*{1.9125cm}\vspace{0.3cm}
\end{center}
\caption{Time sequences of corundum density, corundum grain radius, silicate density, and silicate grain radius for slices through the center of model st28gm06n038 (rows 1--4). The dashed circle marks the estimated stellar radius of 478\,R$_\odot$. The over-plotted contour in the corundum grain size panels (row 2) marks where silicate grains (row 4) have reached a radius of 0.2\,$\mu$m, probably indicating the outer edge of the gravitationally bound dust layers (see text). Row 5 visualizes the dynamics with pseudo-streamlines (integrated over 10$^6$\,s; the colored lines indicate the radial ray used in Fig.~\ref{f:st28gm06n038_1DRay}). 
Shock fronts appear as regions where flows with different directions collide. The snapshots are about 2 months apart. 
\label{f:st28gm06n038_QuSeq2}}
\end{figure*}

\section{Results}\label{s:results}

In this section we give a detailed description of the non-gray model st28gm06n038. The emphasis is on atmospheric dynamics, the dust formation process and the emergence of large-scale structures in the close vicinity of the star. 

\subsection{Convective flows, pulsations, and atmospheric waves}

Global AGB star models are characterized by giant convection cells, which can span 90 degrees or more in cross sections, as discussed in detail in our earlier papers \citep{Freytag2008A&A...483..571F,  Freytag2017A&A...600A.137F}. The cells are outlined by non-stationary downdrafts, which reach from the surface of the convection zone deep down toward the center. While the flow time from the surface to the center is around half a year, the convective cells can have a lifetime of many years. The photospheric layers have a complex, variable appearance due to smaller, more short-lived cells, which form close to the surface of the convection zone. In addition to the convective flows, the 3D RHD models show more or less pronounced radial pulsations, with typical periods of about a year or more, accompanied by variations in luminosity. 

These dynamical processes generate waves of various frequencies and spatial scales, which quickly develop into shock waves as they propagate outward through the atmosphere with its steeply declining density. The shocks give rise to ballistic gas motions, which typically peak around 2 stellar radii. As the shock waves interact and merge, they produce large-scale regions of enhanced densities in their wakes. 

The dynamics of the convective stellar interior and of the atmosphere is illustrated in Fig.\,\ref{f:st28gm06n038_QuSeq1}, showing time series of gas density, radial velocity, entropy, and temperature for slices through the center of the non-gray model (rows 1--4). The surface of the convection zone is characterized by a sharp drop in both entropy and temperature, coinciding with a transition from mostly overturning gas flows in the interior to large-scale shock waves in the atmosphere. A pronounced example of a shock propagating outward through the atmosphere is visible in the upper right quadrant of the velocity plots. In the dense wake of the shock (blue area in the velocity plot, orange in the gas density panels), gas is temporarily lifted to distances where dust formation may occur. The temperature panels feature over-plotted isotherms at 1500 and 1150\,K, roughly corresponding to the expected inner edges of the corundum and silicate dust shells, respectively. It should be noted here that the temperatures in the atmosphere, which are set by non-local radiative processes, show a rather smooth, almost spherical pattern, in contrast to the gas densities, which are strongly affected by the local dynamics. 

Finally, the bottom row of panels in Fig.\,\ref{f:st28gm06n038_QuSeq1} shows the relative surface intensity, illustrating changes of brightness and apparent stellar radius over the course of several months. The variations of photo-center location caused by giant convection cells and their impact on Gaia results for AGB stars have been discussed in detail in a recent paper by \cite{Chiavassa2018A&A...617L...1C}.

\begin{figure}[hbtp]
\includegraphics[width=8.8cm]{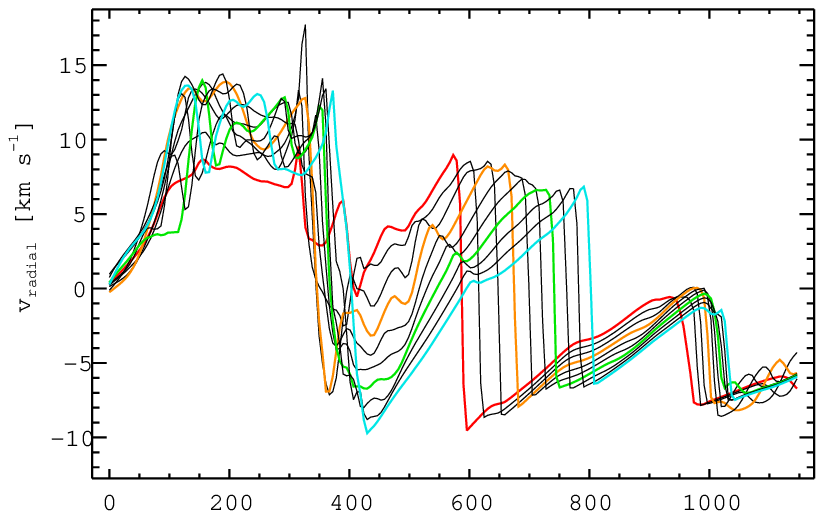}\vspace{-2mm}
\includegraphics[width=8.8cm]{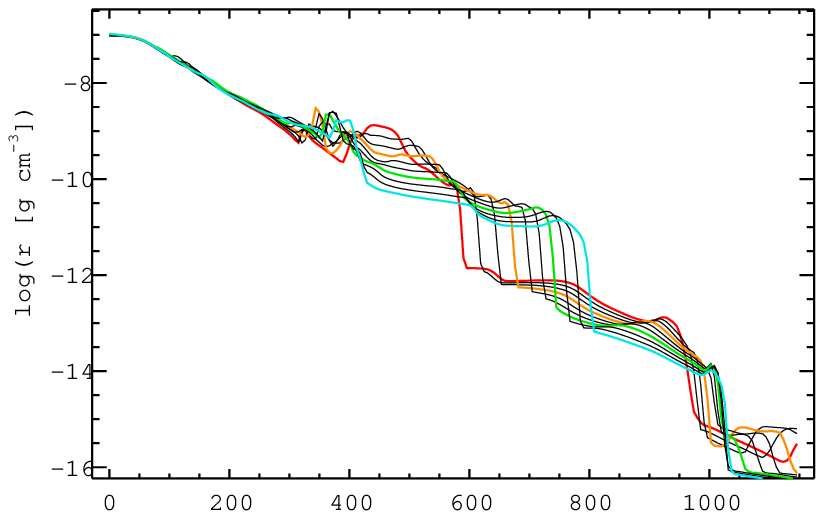}\vspace{-2mm}
\includegraphics[width=8.8cm]{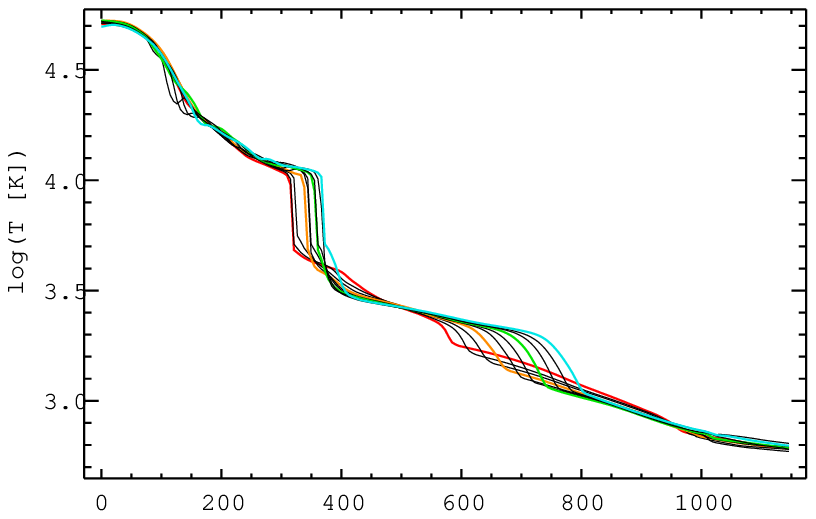}\vspace{-2mm}
\includegraphics[width=8.8cm]{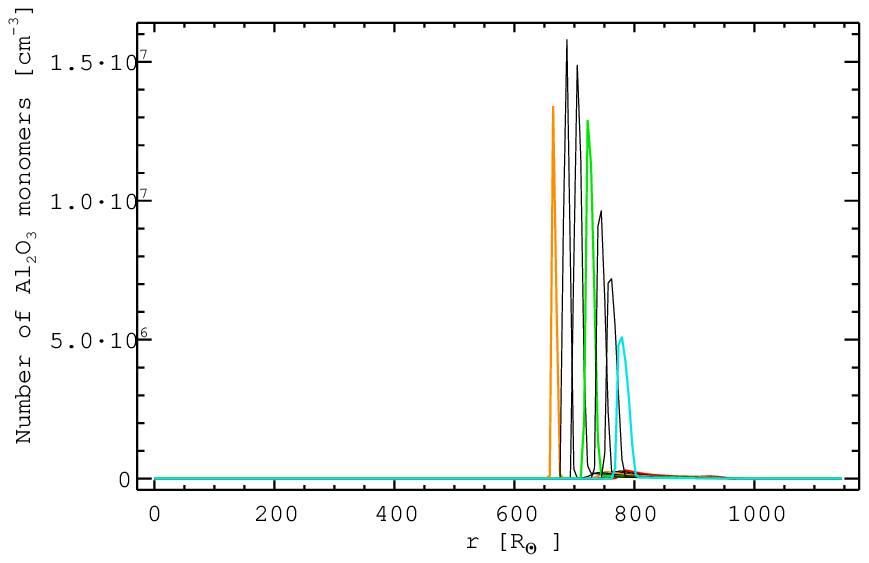}
\caption{One-dimensional radial profiles of velocity, gas density, temperature, and corundum density, showing the formation of a new dust cloud (appearing in the upper right quadrant of Fig.\,\ref{f:st28gm06n038_QuSeq2}). The colors (red, orange, green, blue) mark the four time steps selected in Figs.\,\ref{f:st28gm06n038_QuSeq1} and \ref{f:st28gm06n038_QuSeq2}, the black lines show intermediate snapshots, 1.75$\cdot$10$^6$\,s (20.3\,d) apart. The sharp drop in temperature from 10\,000\,K to about 4000\,K marks the transition from the convection-dominated interior to the shock-dominated atmosphere.
\label{f:st28gm06n038_1DRay}}
\end{figure}

\subsection{Formation of dust clouds}

As outlined in Sect.\,\ref{s:dust_spec}, dust formation is sensitive to atmospheric temperatures and gas densities: Temperature acts as a threshold for the onset of grain growth or evaporation, while gas densities affect the grain growth rates, and therefore the efficiency of dust formation in a dynamical atmosphere. In the following, we demonstrate how the combined effects of temperature and density can produce inhomogeneous dust layers around AGB stars. 

Rows 1--4 of Fig.\,\ref{f:st28gm06n038_QuSeq2} show time sequences of selected dust properties, plotted for slices through the center of the non-gray model, similar to the gas quantities in Fig.\,\ref{f:st28gm06n038_QuSeq1}. The first and third row represent the dust densities (defined as total number of monomers condensed into grains, per volume of atmosphere) for corundum and silicates, respectively. The second and fourth row show the corresponding dust grain radii. To simplify the comparison of patterns in the corundum and silicate distributions, the over-plotted contour in the corundum grain size panels (second row) marks where silicate grains have reached a radius of 0.2 microns. This is a size where radiation pressure should be sufficient to trigger a wind, indicating the possible outer edge of the gravitationally bound dust layers (see below). Row 5 in Fig.\,\ref{f:st28gm06n038_QuSeq2} visualizes the dynamics with pseudo-streamlines (integrated over 10$^6$\,s; the colored lines indicate the radial ray used in Fig.~\ref{f:st28gm06n038_1DRay}). Shock fronts appear as regions where flows with different directions collide. 

In the panels showing the density and grain size of corundum (first and second row of Fig.\,\ref{f:st28gm06n038_QuSeq2}), the formation of a new dust cloud can be traced in the upper right quadrant. The first (leftmost) images of the time sequence show grains from previous dust formation events in the upper right quadrant, located mostly in regions of in-falling material (indicated by red color in the radial velocity plot, Fig.\,\ref{f:st28gm06n038_QuSeq1}, second row). In the next snapshot (second column), an outward-propagating shock wave (marked by a sharp red-blue transition in radial velocity, or a region of colliding flows in Fig.\,\ref{f:st28gm06n038_QuSeq2}, row 5) is just reaching distances where temperatures are low enough for corundum condensation to take place. In the wake of the shock (blue region in the radial velocity plot, Fig.\,\ref{f:st28gm06n038_QuSeq1}, second row), new grains are starting to grow. A small bright spot 
appears in both the corundum density and grain size plots (Fig.\,\ref{f:st28gm06n038_QuSeq2}, second column, row~1 and 2). At the third instance, corundum condensation has occurred over a larger region in the wake of the shock. The dusty gas is moving outward on ballistic trajectories, forming an expanding, crescent-shaped structure. In the fourth snapshot, the new corundum cloud has reached distances where silicate dust can exist (third and fourth row of Fig.\,\ref{f:st28gm06n038_QuSeq2}). Beyond this point, a transition from the gravitationally bound dust layers to a dust-driven outflow can be expected.  

Figure\,\ref{f:st28gm06n038_1DRay} presents the emergence of the new corundum cloud in a form which is more easy to compare to existing spherically symmetric atmosphere and wind models \citep[e.g.][]{Hoefner2016A&A...594A.108H}. The panels show radial structures of velocity, gas density, temperature, and corundum density, along a radial ray from the stellar center outward in the direction of the forming cloud. Each curve plotted represents an instant in the evolution of the structures. The colored lines (red, orange, green, blue) mark the four time steps also shown in Figs.\,\ref{f:st28gm06n038_QuSeq1} and \ref{f:st28gm06n038_QuSeq2}, the black lines correspond to intermediate snapshots. In the radial velocity structures (Fig.\,\ref{f:st28gm06n038_1DRay}, top panel) the propagation of a strong shock front through the atmosphere, from about 600 $R_{\odot}$ to about 800 $R_{\odot}$, is clearly visible, leading to a compression of the gas by about two orders of magnitude (see gas density, second panel). In the first three snapshots (red and two black lines), the temperatures (third panel) in the wake of the shock are still above the condensation limit for corundum. In the next snapshot (orange line), however, the temperature in the compressed gas behind the shock has dropped sufficiently, and a new corundum cloud is starting to form, indicated by the appearance of a spike in the dust density plot (bottom panel). In about a month, the grains grow to their final size (seen as an increase in dust density; next black line), and already two snapshots later the dust density in the cloud is decreasing, similar to the gas density, due to the general expansion of the outward-moving layers.  

When comparing the dust layers plotted in Figs.\,\ref{f:st28gm06n038_QuSeq2} and \ref{f:st28gm06n038_1DRay} to observations or 1D atmosphere and wind models, two points should be kept in mind: First, the dust densities shown in Fig.\,\ref{f:st28gm06n038_QuSeq2} (row~1 and 3) are values for a central slice through the model, not column densities integrated along the line of sight, or synthetic images. In those cases, the superposition of layers along the line of sight would probably make the clouds appear more extended. Examples of synthetic scattered light images based on radial profiles of DARWIN models have been presented by \citet{Aronson2017A&A...603A.116A}, illustrating these effects. Secondly, the linear scale in the dust density plots of the 3D model was chosen to clearly show the emergence of inhomogeneities in the innermost, gravitationally bound dust layers. It emphasizes regions of high density, making the dust layers appear narrow in the radial direction, due to rapidly falling densities with increasing distance from the star. This is in particular true for regions where grain growth is occurring in the dense wakes of shocks, which expand outward into much less dense material. As can be seen in the grain size plots in Fig.\,\ref{f:st28gm06n038_QuSeq2} (row~2 and 4), which are also a measure of the amounts of materials condensed into grains, the dusty layers extend all the way to the outer limits of the computational domain.

\subsection{Overall morphology of the dusty atmosphere}

The development of structures shown in Fig.\,\ref{f:st28gm06n038_1DRay}, tracing the formation of a new dust layer in the wake of a strong shock, that propagates outward through the atmosphere, is reminiscent of results obtained with 1D atmosphere and wind models. However, 1D models assume spherical symmetry, implying that atmospheric structures and dynamics are identical in all radial directions. The large-scale shock fronts propagating through the atmospheres of 3D models, on the other hand, are not spherically symmetric. While a given shock wave may cover a significant fraction of the stellar surface, leading to local conditions in its wake that are comparable to 1D models, the velocities at the shock front are not uniform, and the direction of gas motion is not purely radial. In addition, there are usually several of these large-scale shocks, propagating through the atmosphere in different directions at the same time. Sometimes they collide with each other, creating high density filaments, that appear as bright, almost radial spikes in the density plots in Fig.\,\ref{f:st28gm06n038_QuSeq1}. Occasionally, this results in a corresponding localized structure of strongly enhanced dust density, formed in a process that cannot be treated with 1D simulations. Differences between 1D and 3D models regarding atmospheric dynamics have recently been discussed in detail by \cite{Liljegren2018A&A...619A..47L}.

\begin{figure}[hbtp]

\includegraphics[width=7.04cm]{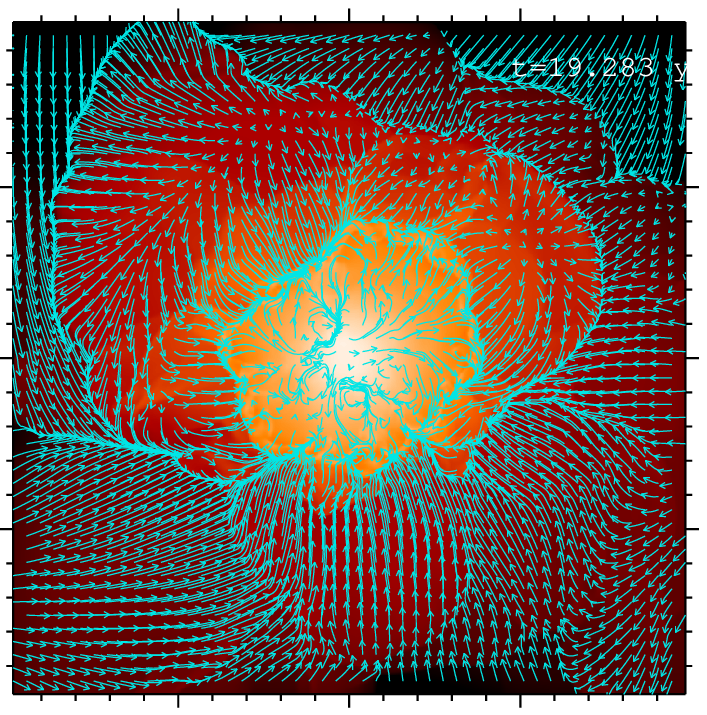}
\includegraphics[width=1.745cm]{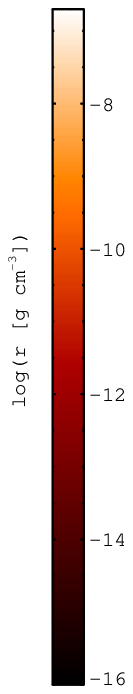}

\includegraphics[width=7.04cm]{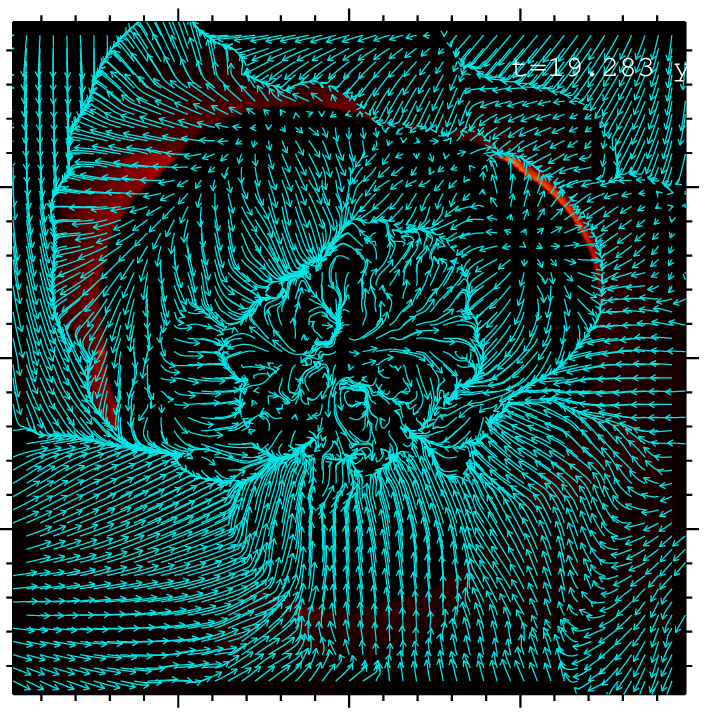}
\includegraphics[width=1.745cm]{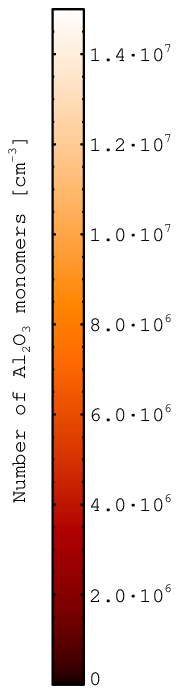}
\vspace{5mm}

\caption{Snapshots of logarithmic gas mass density (top) and corundum density (number density of monomers condensed into grains; bottom) with overplotted pseudo-streamlines integrated over 10$^6$\,s, for a slice through the center of model st28gm06n038, for the same instant as the rightmost column in Figs.\,\ref{f:st28gm06n038_QuSeq1} and \ref{f:st28gm06n038_QuSeq2}.
\label{f:st28gm06n038_rhoAndStreamlines}}

\end{figure}

\begin{figure}[hbtp]

\includegraphics[width=7.04cm]{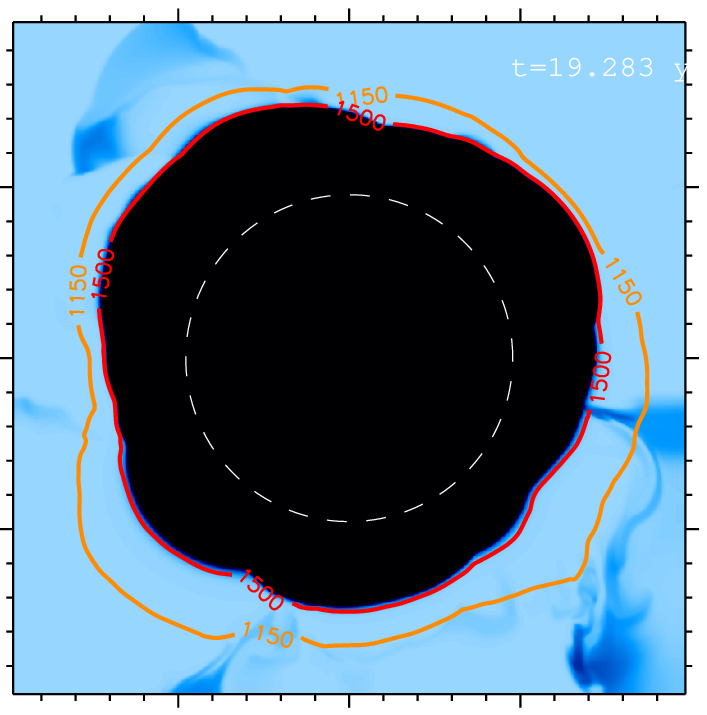}
\includegraphics[width=1.745cm]{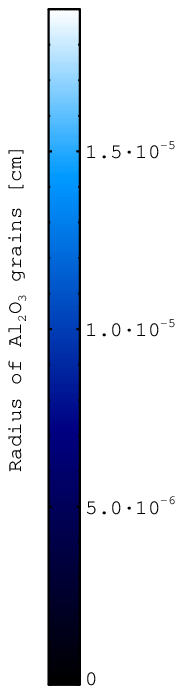}

\includegraphics[width=7.04cm]{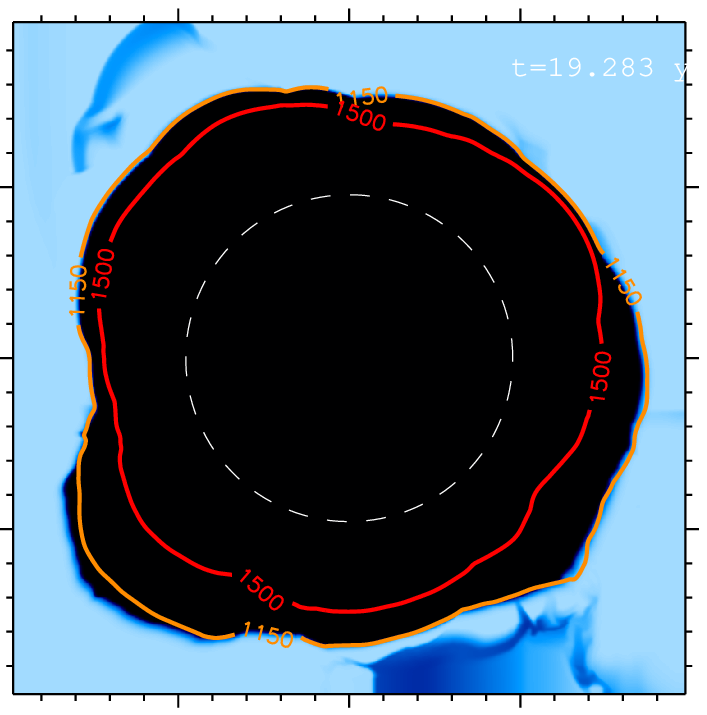}
\includegraphics[width=1.745cm]{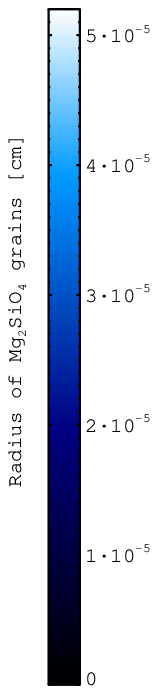}
\vspace{5mm}

\caption{Snapshots 
of corundum grain radius (top) and silicate grain radius (bottom), with over-plotted isotherms at 1500\,K and 1150\,K (same as in Fig.\,\ref{f:st28gm06n038_QuSeq1}, row 4), for a slice through the center of model st28gm06n038, at the same instant as in Fig.\,\ref{f:st28gm06n038_rhoAndStreamlines}. The dashed circle marks the average stellar radius. 
\label{f:st28gm06n038_RdustTandrhoContour}}

\end{figure}

The non-spherical shock fronts in the 3D models, triggered by convection and pulsations, lead to a patchy distribution of dust in the atmosphere. Grain growth is more efficient in high-density regions, and new dust is therefore concentrated in the wakes of outward-propagating shocks, apparent as arc-like structures in the central cross-sections shown in Fig.\,\ref{f:st28gm06n038_QuSeq2} (row~1 and 3). Seen face-on, these structures correspond to partial dust cloud layers, covering an area of the stellar surface similar to the shock. 

Figure\,\ref{f:st28gm06n038_rhoAndStreamlines} illustrates the interplay between dynamics, gas densities, and dust condensation. The upper panel shows gas density with overplotted pseudo-streamlines for a central slice of the non-gray model, while the lower panel combines a plot of corundum density with pseudo-streamlines for the same snapshot. The shock fronts, apparent as regions where flows with different directions collide, correspond to abrupt changes in gas density (upper panel). The newly-formed dust cloud in the upper right quadrant (lower panel) is located in the dense wake of the outward-propagating shock, as described above. A second dust cloud, visible in the upper left quadrant, is the result of a separate dust formation event. Overall, like the gas density, the dust density decreases strongly with increasing distance from the star, resulting in a limited radial range of observable structures. 

In contrast to the dust density distributions, the grain radii plots (Fig.\,\ref{f:st28gm06n038_QuSeq2}, rows~2 and 4; Fig.\,\ref{f:st28gm06n038_RdustTandrhoContour}) do no simply reflect the current atmospheric structures and dynamics (density, velocity, location of shock fronts, etc.). This is illustrated in more detail by a comparison of Figs.\,\ref{f:st28gm06n038_rhoAndStreamlines} and \ref{f:st28gm06n038_RdustTandrhoContour}. The complex spatial patterns in grain sizes result from local grain growth rates, that vary during the condensation process, spanning timescales that are comparable to those of atmospheric dynamics (typically weeks to months). In other words, dust grain properties are not set instantaneously; instead they preserve, to some degree, a record of the changing ambient conditions during the condensation process. 

The inner edges of the corundum and silicate layers are largely defined by temperature, acting as a threshold for the onset of grain growth and evaporation. We note, again, that the atmospheric temperature distribution (Fig.\,\ref{f:st28gm06n038_QuSeq1}, row 4), is set by non-local radiative processes and tends to be more smooth and quasi-spherical than the densities, which are dominated by local dynamics and shocks. As shown in Fig.\,\ref{f:st28gm06n038_RdustTandrhoContour}, the inner edges of the corundum and silicate dust shells correspond almost perfectly to isotherms. Taking a closer look at the corundum grain sizes (upper panel), however, some small, but systematic, deviations can be detected: In the lower right quadrant, gas and dust are falling toward the star (see streamlines in Fig.\,\ref{f:st28gm06n038_rhoAndStreamlines}), and the grains start to shrink due to thermal evaporation when temperatures exceed a critical value (see Sect.\,\ref{s:dust_spec}). Since the dust does not evaporate instantaneously, the blue region extends a little bit beyond the 1500\,K isotherm to higher temperatures. The opposite situation is seen in the upper left quadrant. There, the flow is directed outward toward regions of lower temperature, but it takes time for the grains to grow and reach their full size. Consequently, the blue zone starts somewhat  outside the 1500\,K contour line. In summary, the deviations of the inner edges of the corundum and silicate layers from isotherms demonstrate the time-dependent, non-equilibrium nature of the dust condensation and evaporation processes. The effects on the dust distribution will tend to be stronger at lower densities, and they also depend on the flow velocities. For the edge of the silicate shell, some more pronounced deviations from the 1150\,K isotherm, located further out at lower densities, can be noticed in the lower panel of Fig.\,\ref{f:st28gm06n038_RdustTandrhoContour}. 

\begin{figure*}[hbtp]
\begin{center}

\includegraphics[width=16.0cm]{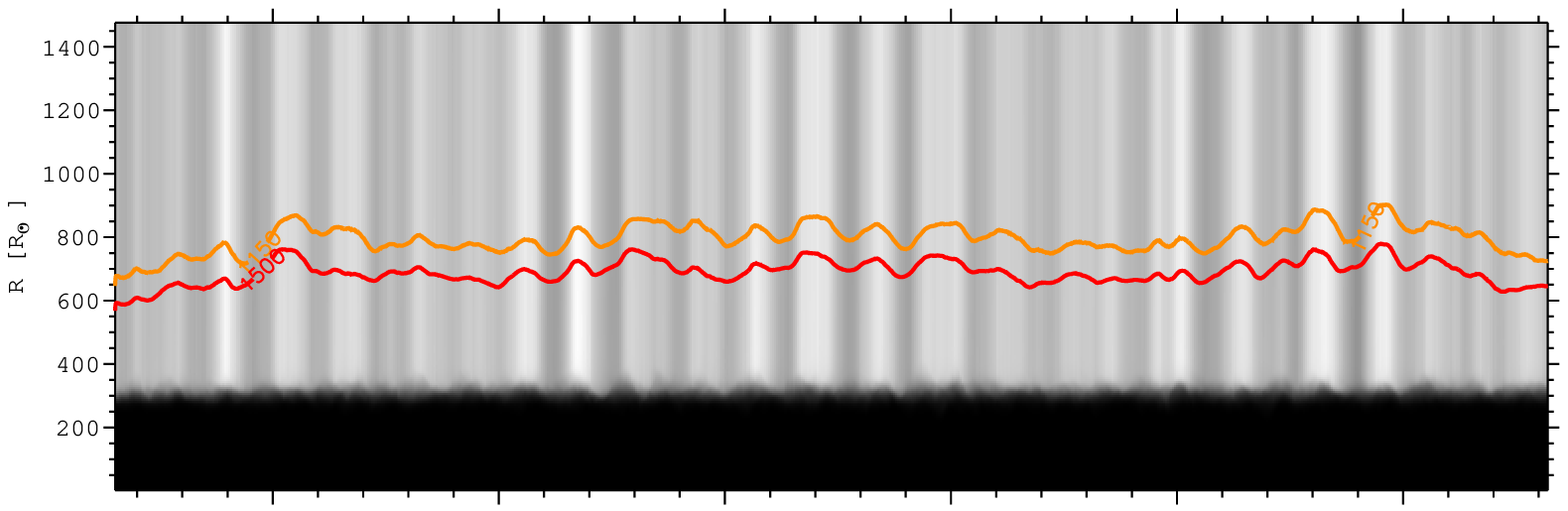}\includegraphics[width=2.0cm]{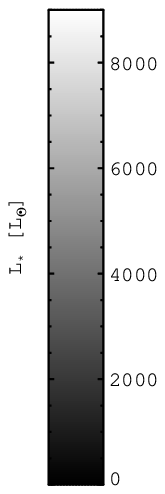}

\includegraphics[width=16.0cm]{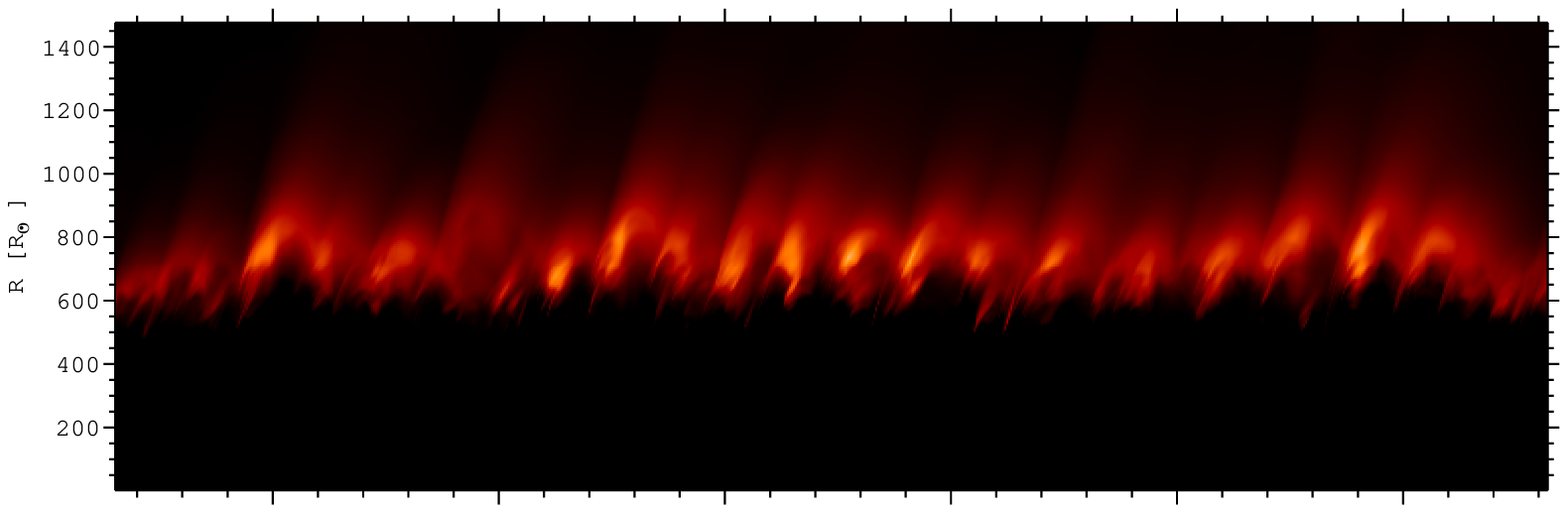}\includegraphics[width=2.0cm]{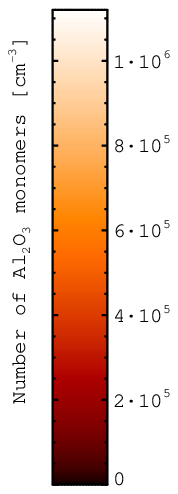}

\includegraphics[width=16.0cm]{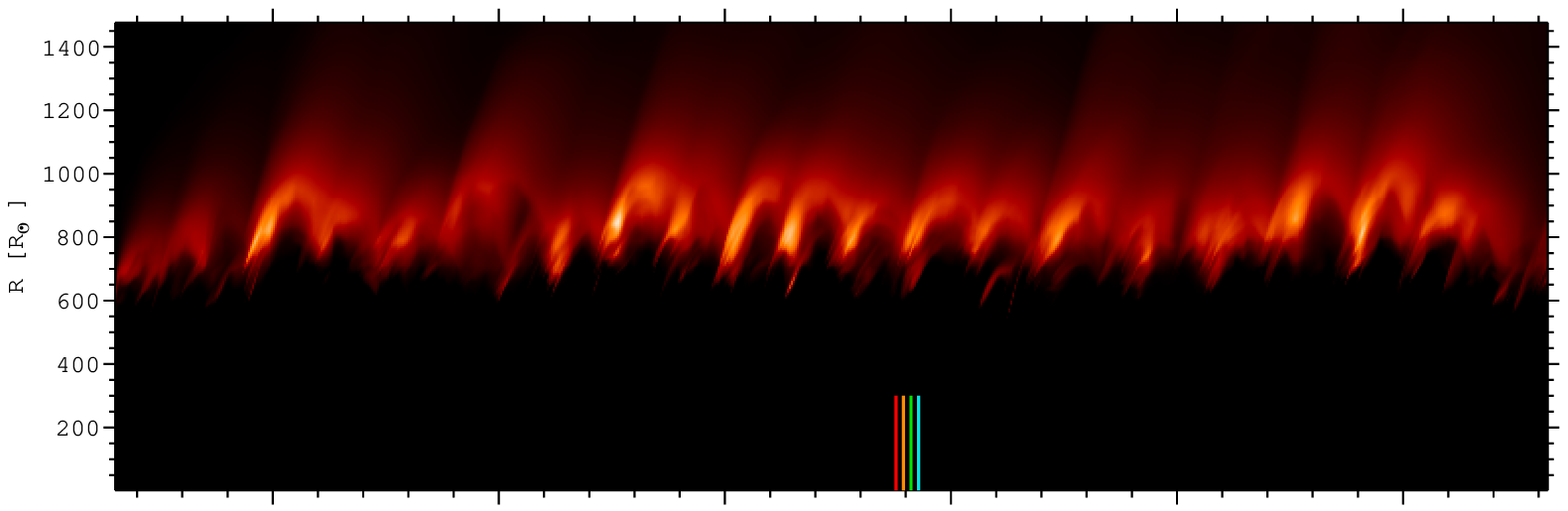}\includegraphics[width=2.0cm]{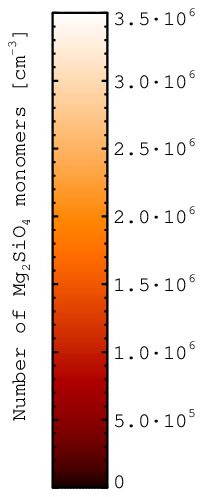}\vspace{0.4cm}
\end{center}
\caption{Spherical averages of luminosity (integrated radiative flux), corundum density, and silicate density in model st28gm06n038, as a function of radial distance and time. In the top panel, two isotherms with temperatures representative of corundum and silicate condensation are shown (at 1500\,K and 1150\,K, respectively, same as in Figs.\,\ref{f:st28gm06n038_QuSeq1} and \ref{f:st28gm06n038_RdustTandrhoContour}), to demonstrate the interplay between variable radiative heating and the inner edges of dust layers. The dark area in the lower part of the top panel represents the deep stellar interior, where most of the energy flux is transported by convection. The colored vertical lines in the bottom panel mark the instants selected for Figs.\,\ref{f:st28gm06n038_QuSeq1} and \ref{f:st28gm06n038_QuSeq2}.
\label{f:st28gm06n038_QuOvertimeAndx}}

\end{figure*}

\subsection{Time-dependence of global quantities}

Long-period variability of global properties like luminosity and effective temperature is a characteristic feature of AGB stars, which realistic dynamical models need to reproduce. The large-amplitude variations observed in Mira stars are commonly attributed to radial pulsations. To extract information about radial pulsations from 3D models with their complex large-scale convective flows, \citet{Freytag2017A&A...600A.137F} used averages of radial velocities and other quantities, taken over spherical shells. An analysis of the variations with time and depth revealed radial pulsations with more or less pronounced dominant frequencies, that are in general agreement with the observed period-luminosity relation of \cite{Whitelock2009MNRAS.394..795W}. 

Here, we are interested in a similar global view of the dusty atmospheric layers and their variability, in order to facilitate a comparison with 1D dynamical models and unresolved observations (photometry, spectroscopy). Figure\,\ref{f:st28gm06n038_QuOvertimeAndx} shows spherical averages of luminosity (frequency-integrated radiative flux; top panel), corundum density (middle), and silicate density (bottom panel), changing with radial distance and time. 

In the top panel of Fig.\,\ref{f:st28gm06n038_QuOvertimeAndx} (showing radiative luminosity), the dark area in the lower part, representing the deep stellar interior, indicates where most of the energy flux is transported by convection. Further out, in the stellar atmosphere, the energy is mostly carried by the radiative flux, which shows pronounced variations on timescales of about a year, consistent with the radial pulsations. Atmospheric temperatures are mostly set by radiative processes, dominated by photons emitted from the stellar surface. The strong coupling between radiative flux and temperature is reflected by two over-plotted isothermal lines (same temperatures as in Figs.\,\ref{f:st28gm06n038_QuSeq1} and \ref{f:st28gm06n038_RdustTandrhoContour}, indicating thresholds for the formation and destruction of corundum and silicate dust). Tracing the depth of layers with given temperature, the lines move outward for increasing luminosity and inward for decreasing luminosity, reflecting changing levels of radiative heating.  

The relatively well-defined mean inner edges of the corundum and silicate shells (middle and bottom panels of Fig.\,\ref{f:st28gm06n038_QuOvertimeAndx}) follow the same trend as the isothermal lines in the top panel. This is expected, since the onset of dust formation and destruction is largely defined by temperature (see Fig.\,\ref{f:st28gm06n038_RdustTandrhoContour}). Further out, the variations in the dust densities show indications of outward-propagating shock waves, that is, high-density areas moving upward as time progresses, representing the wakes of the shocks. This picture is consistent with the spherical averages of radial velocities studied in \citet{Freytag2017A&A...600A.137F}, and reminiscent of 1D dynamical atmosphere models. The mean radial dynamics of the 3D model follows qualitatively similar patterns as the motions of spherical mass shells in 1D models \citep[see, e.g.,][]{Liljegren2017A&A...606A...6L, Liljegren2018A&A...619A..47L}. However, while there is a clear periodicity in the spherically-averaged properties the 3D model, related to the radial pulsation, they also show pronounced cycle-to-cycle variations, reflecting the large-scale asymmetries in the dynamical atmosphere.

\section{Discussion: Models and observations}\label{s:discussion}

Observational evidence for non-spherical dynamical structures in atmospheres of AGB stars has been accumulating over the past few decades, following improvements in high-angular resolution techniques. In recent years, direct imaging in the visual and near-IR, as well as image reconstruction from IR interferometric data, have given an increasingly more detailed picture of the dynamical atmospheres and, in particular, the dust-forming layers. 

An object, that has been observed frequently using various techniques, is the nearby C-type AGB star IRC+10216 (CW~Leo). Recently, \cite{Stewart2016MNRAS.455.3102S} studied the dynamical evolution of dust clouds, using images reconstructed from aperture-masking interferometric observations obtained with Keck and VLT, and from occultation measurements by the Cassini spacecraft. They find changes on timescales of years, and a completely different circumstellar morphology compared to observations by \citet{Kastner1994ApJ...434..719K}, \citet{Haniff1998A&A...334L...5H} and \citet{Weigelt1998A&A...333L..51W} about 20 years earlier. 
In this star, the stellar surface and atmosphere are obscured by the optically thick envelope of carbon dust, making it difficult to relate the observed structures to processes in the dust-forming layers. Even in carbon stars with more moderate mass loss, however, IR images can be difficult to interpret, as demonstrated by the recent example of R Scl, see \citet{Wittkowski2017A&A...601A...3W}. 

The usually more transparent dusty envelopes of M-type and S-type AGB stars, on the other hand, give better access to the stellar atmospheres, improving the possibilities to identify the physical mechanisms that produce inhomogeneous dust layers. \citet[][]{Paladini2018Natur.553..310P} recently presented H-band images of the star {$\pi$}$^{1}$ Gruis, reconstructed from VLTI/PIONIER data, which show evidence of large granulation cells on the stellar surface. The sizes of the observed surface structures agree well with extrapolations of local 3D models for less evolved stars, and with qualitative predictions of global 3D AGB star models. As discussed in Sect.\,\ref{s:results}, the large-scale convective flows result in a strongly dynamical, inhomogeneous atmosphere, which, in turn should lead to the formation of clumpy dust clouds in the close stellar environment. Such dust clouds have recently been detected in scattered light observations of R~Dor and W~Hya in the visual and near-IR range with VLT/SPHERE-ZIMPOL, showing changes in morphology on timescales of weeks to months \citep[][]{Khouri2016A&A...591A..70K, Ohnaka2017A&A...597A..20O}, as well as changes in dominant grain size, from 0.1\,$\mu$m at minimum light to 0.5\,$\mu$m at pre-maximum light in the SRa variable W\,Hya \citep{Ohnaka2016A&A...589A..91O, Ohnaka2017A&A...597A..20O}. The dynamical timescales implied by changes in circumstellar morphology, the timescales of grain growth, and the grain sizes, are in good agreement with the 3D model presented here and DARWIN models for M-type AGB stars \citep[][]{Hoefner2008A&A...491L...1H, Bladh2013A&A...553A..20B, Bladh2015A&A...575A.105B, Hoefner2016A&A...594A.108H, Aronson2017A&A...603A.116A, Liljegren2017A&A...606A...6L}. The observed spatial scales, that is, the sizes of clouds and the distances from the stellar surface, compare well with the 3D model results. Recent sub-mm continuum and line observations of W Hya with ALMA \citep[][]{Vlemmings2017NatAs...1..848V} give further evidence of a dynamical atmosphere with a complex morphology, featuring a hot spot presumably due to local shock heating, as well as extended inhomogeneous layers of warm and cool gas. The cool gas component (about 900\,K) can be traced out to around 2.5 stellar radii, reaching well into the dust formation zone. It shows both infall and outflow velocities, similar to our 3D models. 

In summary, the picture presented in Sect.\,\ref{s:results}  is in good agreement with recent high-angular resolution observations. This indicates that the formation of inhomogeneous dust layers due to large-scale atmospheric shocks, triggered by convection and pulsations, is a likely scenario. This mechanism operates in the innermost, gravitationally bound region of the dust envelope, which is probably dominated by corundum grains \cite[see, e.g.,][]{Karovicova2013A&A...560A..75K, Khouri2015A&A...577A.114K, Hoefner2016A&A...594A.108H}, before an outflow is triggered by radiation pressure on silicate dust. 

At present, one can only speculate how the onset of a stellar wind would affect the results, since no global 3D RHD models of AGB stars with dust-driven outflows exist so far. Since the basic cloud-formation mechanism seems to precede wind acceleration, it should not be changed qualitatively by the presence of a stellar wind. Judging from DARWIN models, a likely quantitative difference could be a lower fraction of silicon condensation, compared to the almost complete condensation found in model st28gm06n038. When the dust grains reach sizes where the outward-directed radiative force exceeds stellar gravity, triggering an outflow, condensation slows down drastically due to rapidly decreasing densities. As a consequence, significant fractions of the condensible chemical elements may remain in the molecular gas. Typical condensation degrees of Si in DARWIN models are about 20--30\,\% \citep[][]{Bladh2015A&A...575A.105B, Hoefner2016A&A...594A.108H}.

\section{Conclusions}\label{s:conclusions}

In this paper, we have presented new 3D radiation-hydrodynamical simulations, performed with the CO5BOLD code, which explore the formation of clumpy dust clouds in M-type AGB stars. The models cover the outer convective envelope and the atmosphere of the star, including the dust formation region. They account for frequency-dependent gas opacities and include a time-dependent description of grain growth and evaporation for corundum and olivine-type silicates.

The dust formation process in atmospheres of AGB stars is sensitive to ambient temperatures and gas densities, which both leave observable imprints: 
\begin{itemize} 
\item
Atmospheric temperatures are mostly determined by the strongly variable radiative flux emitted by the pulsating star. Temperature acts as a threshold for the onset of dust condensation and evaporation, thereby defining the inner edge of the dusty envelope, which is roughly spherically symmetric, in contrast to the dust distribution within the envelope.
\item 
Gas densities are strongly dependent on atmospheric dynamics and, in particular, on shock waves triggered by large-scale convective flows and pulsations. They affect the grain growth rates and, consequently, the efficiency of dust condensation. The non-spherical shock fronts in the 3D models lead to a patchy distribution of dust in the atmosphere, similar to recent high-angular-resolution images. Grain growth is mostly occurring in the wakes of outward-propagating shocks. 
\end{itemize}
The mechanism of cloud formation primarily operates at the inner edge of the dusty envelope, which is dominated by corundum (Al$_2$O$_3$). The formation of silicate dust (Mg$_2$SiO$_4$), at somewhat larger distances from the star, probably indicates the outer limit of the gravitationally bound layers, where radiation pressure may trigger a stellar wind. In this context, it should be noted that the first exploratory 3D models of dust cloud formation in M-type AGB stars presented here do not account for radiative pressure, which would require a computational effort well beyond the scope of this paper. However, with cloud formation preceding wind acceleration, the basic scenario 
described here should also hold in the presence of an outflow. 

In addition to the 3D structures emerging in the models, we have also analyzed spherical means of dust densities and other quantities, which can be more easily compared to unresolved observations and 1D atmosphere and wind models. These quantities show a pronounced periodicity, related to radial pulsations and variable radiative heating of the atmosphere, as well as cycle-to-cycle variations. The latter reflect both, a range of ballistic timescales in the atmosphere, similar to results of 1D models, and the variable 3D morphology of the dusty envelope. Observed cycle-to-cycle variations in photometric and spectroscopic data hold information on both, radial dynamics and non-spherical effects, but monitoring of nearby objects with high-angular-resolution techniques is essential for disentangling them.

\begin{acknowledgements}

This work has been supported by the Swedish Research Council (Vetenskapsr{\aa}det). The computations were performed on resources (``rackham'') provided by SNIC through Uppsala Multidisciplinary Center for Advanced Computational Science (UPPMAX) under Projects snic2017-1-41 and snic2018-3-74. We thank Bernhard Aringer for providing input data for the frequency-dependent gas opacity tables used in the new 3D models, Sofia Ram\-stedt for helpful comments concerning the manuscript, and the referee, Jan Martin Winters, for constructive suggestions.
 
\end{acknowledgements}

\bibliographystyle{aa}    
\bibliography{aa_redsg}

\end{document}